# Observation of Indirect Ionization of $W^{7+}$ in EBIT plasma

Q. Lu[1,2], J. He[1,2], H. Tian[1,2], M. Li[1,2], Y. Yang[1,2], K. Yao[1,2], C. Chen[1,2], J. Xiao[1,2,*], J.G. Li[3,†], B. Tu[4,‡] and Y. Zou[1,2]

[1] Institute of Modern Physics, Department of Nuclear Science and Technology, Fudan University, Shanghai 200433, China

[2] Key Laboratory of Nuclear Physics and Ion-beam Application (MOE), Fudan University, Shanghai 200433, China

[3] Institute of Applied Physics and Computational Mathematics, Beijing 100088, China

[4] Max-Planck-Institute für Kernphysik, Saupfercheckweg 1, 69117 Heidelberg, Germany

Corresponding author: [*]xiao_jun@fudan.edu.cn, [†]Li_Jiguang@iapcm.ac.cn, [‡]bingsheng.tu@mpi-hd.mpg.de

In this work, visible and EUV spectra of $W^{7+}$ have been measured using the high-temperature superconducting electron-beam ion trap at the Shanghai EBIT laboratory under extremely low-energy conditions (lower than the nominal electron beam energy of 130 eV). The relevant atomic structure has been calculated by using the flexible atomic code package based on the relativistic configuration interaction method. The GRASP2K code in the framework of the multi-configuration Dirac-Hartree-Fock method is employed as well when high-precision atomic parameters are required. The $W^{7+}$ spectra are observed 2 charge states in advance according to the ionization energy of $W^{6+}$. A hypothesis for the charge-state evolution of $W^{7+}$ is proposed based on our experimental and theoretical results, that is, the occurrence of $W^{7+}$ ions results from indirect ionization caused by cascade excitation between some metastable states of lower-charge-state W ions, at the nominal electron beam energy of 59 eV.

## I. Introduction

As a metal with the highest melting point, tungsten is considered to be the optimal candidate for wall material of divertor in Tokamak because of its numerous superb properties [1, 2]. However, plasma-wall interaction would make tungsten pass into the core plasma as impurities, which may finally lead to the flameout of fusion [3]. On the other hand, radiation from tungsten ions could carry information about the plasma state, and thus it is essential to obtain and analyze the spectra of tungsten. Since electron beam ion traps (EBITs) employ quasi-monoenergetic and energy-adjustable electron beam to ionize trapped ions, capable of providing specific ions with any targeted charge state, it has been proved to be a good tool for use in disentanglement studies of atomic processes in plasmas during the recent years [4].

So far, a lot of studies have been carried out for the highly charged tungsten ions related to the core plasma in Tokamak since corresponding atomic systems are relatively simple [5-17]. With respect to lowly charged tungsten ions ($W^+$-$W^{13+}$) existing in the boundary plasma, their more complex atomic structures due to the number of electrons, especially the open 4$f$ subshell and competition of orbital energies between 4$f$ and 5$p$ electrons, result in the difficulty in theoretical calculation [18], and furthermore in line identification.

With development of the low-energy EBIT, some progresses have been made on the atomic spectra for lowly charged tungsten ions. For example, spectra of $W^{11+}$-$W^{15+}$ in the 17-26 nm region were measured and analyzed by Li *et al*. [19]. Moreover, Li *et al*. found a strong visible line from $W^{11+}$ [20]. Experiments on $W^{13+}$ were conducted by different EBIT groups as well [21, 22]. For $W^{8+}$-$W^{12+}$ ions, however, spectral data is still rare. In addition to EBIT plasma, a lot of works on lowly charged tungsten ions have been done in vacuum spark plasma [23-30].

It can be seen from Ref. [31] that the ionization energy of $W^{6+}$ ions (122.01±0.06 eV) is much larger than that of $W^{5+}$ ions (64.77±0.04 eV). The opening of the 4$f$ subshell ($4f^{13}5s^25p^6$) may account for this big gap in the ionization energy, and has attracted extensive attention to $W^{7+}$ ions. For example, experiments on $W^{5+}$-$W^{7+}$ in EUV range was conducted by Livermore EBIT Laboratory [32]. Mita *et al*. reported their direct observation of the M1 transition between the fine structure belonging to the ground configuration of $W^{7+}$ ions [33]. According to their results, the M1 line appeared in advance compared with theoretical ionization energy of 122 eV. Therefore, Mita *et al*. proposed that the occurrence of $W^{7+}$ may arise from ionization through the metastable excited states of lower charged tungsten ions. However, this hypothesis has not been confirmed yet.

As for this indirect ionization process, there exists some relevant reports. For example, the occurrence of $Sn^{11+}$-$Sn^{14+}$ below ionization energy was found by Windberger *et al*. [34]. Sakoda *et al*. proposed that $Ba^{11+}$ could appear earlier than expected through indirect ionization from the metastable state of $Ba^{10+}$ [35]. Moreover, Qiu *et al*. discovered some excited metastable states with extraordinarily high population in $W^{28+}$ [36]. However, the specific study on charge-state evolution with indirect ionization has not been carried out until now.

In this work, the spectra of $W^{7+}$ ions in visible and EUV range are measured at the high-temperature superconducting electron-beam ion trap (SH-HtscEBIT) [37]. The atomic structures of $W^{5+}$, $W^{6+}$ and $W^{7+}$ are calculated using the relativistic configuration interaction (RCI) method implemented in the flexible atomic code (FAC) package [38, 39]. In addition, GRASP2K code [40, 41], based on the multi-configuration Dirac-Hartree-Fock theory, is also employed to calculate the energy structure when high-precision atomic data are required.

**II. Experimental setup**

SH-HtscEBIT is utilized for experimental part in this work, whose main structure contains electron gun, drift tube (including DT1, DT2 and DT3), superconducting coil and collector. The special electromagnetic optical structure enables the minimum electron energy of SH-HtscEBIT to reach merely 30 eV, which is extremely suitable for the study of boundary plasma related lowly charged tungsten ions [19, 37].

Tungsten atoms are injected into the trap using W(CO)$_6$ gas with an injection pressure of $1.0 \times 10^{-6}$ torr while the trap vacuum maintains $1.0 \times 10^{-9}$ torr. Once tungsten ions are generated in the central drift tube trap region, they are confined axially by the potential well (100 V) while radially by the space charge effect of electrons and the magnetic field (0.2 T). The trapped ions are collided with the electron beam which is accelerated by the potential difference between DT2 and cathode. Finally, photon radiation from excited states is detected by an Andor Shamrock 303 spectrometer for visible range and a grazing incidence flat-field spectrometer for EUV range [42] respectively.

**III. Theoretical calculation**

An integrated software package FAC is used in this work, which can produce atomic structure, such as energy levels, transition rates, collision (de)excitation rates and so on [14, 38, 39].

In order to simulate spectra under different plasma conditions, a collisional-radiative model (CRM) implemented in FAC is adopted [43, 44]. Here, a balanced system is established in CRM to obtain the energy level population. In the environment of the low-energy EBIT, three main dynamic processes involving electron collision excitation, electron collision de-excitation and radiation decay are included, while other processes like charge exchange and radiation recombination are ignored. On the basis of this assumption, the differential rate of the population of each energy level can be expressed as

$$\frac{dN_i}{dt} = \sum_{j>i}(A^r_{j\to i} \cdot N_j) + \sum_{j<i}(C^e_{j\to i} \cdot N_j) + \sum_{j>i}(C^d_{j\to i} \cdot N_j)$$

$$- \sum_{j<i}(A^r_{i\to j} \cdot N_i) - \sum_{j>i}(C^e_{i\to j} \cdot N_i) - \sum_{j<i}(C^d_{i\to j} \cdot N_i),$$

where $N$ is the population number, the subscripts $(i, j)$ represent the initial or the final energy levels, and $A^r, C^e, C^d$ stand for the radiation decay rate, the electron collision excitation rate, and the electron collision de-excitation rate, respectively. Considering equilibrium condition $\frac{dN_i}{dt} = 0$ and normalized condition $\sum_i N_i = 1$, we can solve the equation above and further obtain the population of each energy level.

The line intensity can be calculated, once level populations and transition rates are given. The simulated spectra are presented with wavelength (given by RCI) and intensity (given by CRM) for analyzing the experimental spectra.

**IV. Results and discussion**

**A. Visible line of $W^{7+}$**

Spectra in the range of 559-623 nm from tungsten ions, which are obtained at the nominal electron beam energy of 55, 58, 59, 70, 90 and 130 eV are shown in Fig. 1.

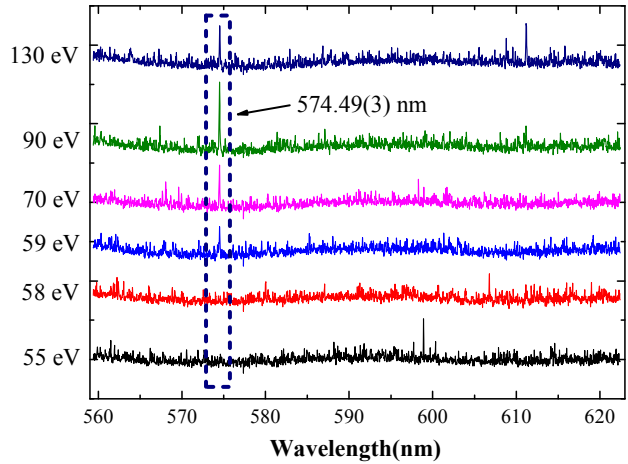

FIG. 1. Spectra of tungsten obtained by SH-HtscEBIT at the nominal electron beam energy of 55, 58, 59, 70, 90 and 130 eV in the range of 559-623 nm. The accumulation time of each spectrum is 2 hours. The line at 574.49(3) nm is the M1 transition between the fine structure splitting in the $4f^{13}5s^25p^6$ $^2F$ ground term of $W^{7+}$.

Line at 574.49(3) nm just appears when the nominal electron beam energy is tuned from 58 eV to 59 eV, indicating that a new charge state is created. We also find a dependence of the line intensity on the electron beam energy, which becomes maximum at nearly 90 eV, and decreases as the energy is at 130 eV.

Since the nominal electron beam energy only represents the voltage difference between the cathode and the central drift tube DT2, the real electron beam energy must be corrected from that. Usually the electron beam energy can be given in the following expression [45]

$$E_e(\text{eV}) = e \cdot (V_{\text{DT2}}(V) - V_{\text{Cathode}}(V) + V_{\text{sp}}(V)),$$

where $E_e$ is the electron beam energy, $V_{\text{DT2}}$ the voltage of DT2, $V_{\text{Cathode}}$ the voltage of cathode and $V_{\text{sp}}$ the potential produced by space charges.

The correction of the electron beam energy is divided into two parts. The first part is the power supply correction, which is a deviation between the set value and the output value of the power supplies. One multimeter (Fluke 17B) is used to measure the actual output voltage and the results are listed in Table I.

The second part is the correction from the space charge effect. The space charge effect $V_{\text{sp}}$, which is typically several tens of eV, results in the reduction of electron beam energy. In case of lowly charged tungsten ions, the ionization energy interval of adjacent charged ions is comparable to $V_{\text{sp}}$, and thus confuse the charge state identification. The space charge effect can be estimated by [46]

$$V_{\text{sp},n}[V] = \frac{30 I_e[A]}{\sqrt{1-\left(\frac{E_{\text{set}}-eV_{\text{sp},n-1}}{511000}[\text{eV}]+1\right)^{-2}}} (\ln\left(\left(\frac{r_e}{r_{\text{dt}}}\right)^2-1\right).$$

In the equation above, $I_e$ (2-3 mA in this case) represents the value of electron beam current; $E_{\text{set}}$ is the potential difference between the DT2 and cathode; $r_e$ stands for the radius of electron beam, typically 150 μm; and $r_{\text{dt}}$, 1 mm, labels the radius of drift tube.

In addition to electrons, ions also have a space charge effect, which compensates for the influence of electrons. Here a coefficient of 0.4 is introduced based on the results in Ref. [46], where the experimental conditions are very similar to ours. It should be noticed that this coefficient may introduce an uncertainty of about 10% in this case. The corrected electron beam energy and the uncertainties are displayed in Table I. The ionization energy of tungsten is also listed in Table II.

TABLE I. Correction of the electron beam energy: the set potential difference between the cathode and DT2 (nominal electron beam energy) $E_{set}$, the output potential difference between the cathode and DT2 $E_{out}$, the space charge effect from electrons and ions $V_{sp}$, and the finally corrected electron beam energy $E_{corr}$. The uncertainty for $V_{sp}$ and $E_{corr}$ are also given.

| $E_{set}$ (eV) | $E_{out}$ (eV) | $V_{sp}$ (eV) | $E_{corr}$ (eV) |
|---|---|---|---|
| 58.0 | 64.4 | 13.4±2.5 | 51.0±2.5 |
| 59.0 | 65.4 | 10.5±2.1 | 54.9±2.1 |

TABLE II. Ionization energy of tungsten [31].

| Ion charge | Ionization energy (eV) |
|---|---|
| +3 | 38.2±0.4 |
| +4 | 51.6±0.3 |
| +5 | 64.77±0.04 |
| +6 | 122.01±0.06 |

Based on the relation between the corrected electron beam energy and tungsten ionization energy, the line at 574.49 nm appears as long as the electron beam energy exceed the ionization energy of $W^{4+}$, i.e. 51.6 eV, rather than $W^{6+}$. The experimental results indicate that the line at 574.49 nm could not come from $W^{7+}$, while from those of charge states under 7+. To identify this line, the RCI method in the FAC package is used to calculate the atomic structure of the $W^{5+}$, $W^{6+}$ and $W^{7+}$. Part of their energy levels is shown in Fig. 2, Fig. 3 and Fig. 4 respectively.

According to the calculations, lines from $W^{5+}$ ions with strong intensity are not in visible range, but in the EUV range instead (see Table III). The strong M1 transition line $^2D_{5/2} - ^2D_{3/2}$ in the ground configuration $4f^{14}5s^25p^65d^1$ lies in the infrared range.

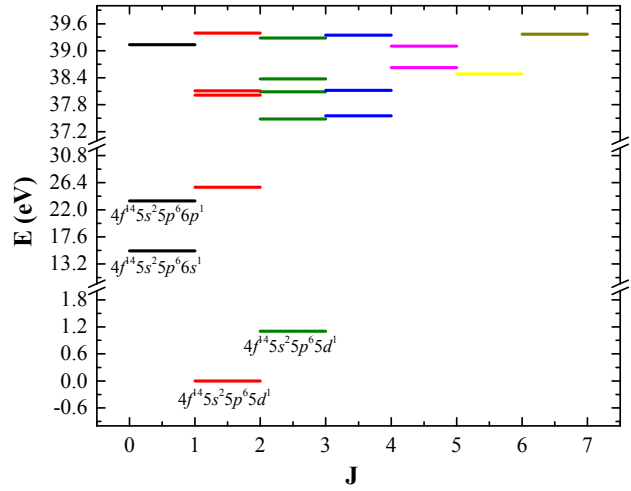

FIG. 2. A partial energy level diagram of $W^{5+}$ with the lowest, in energy, 20 energy levels from FAC calculations.

TABLE III. Results of CRM for $W^{5+}$ ions by FAC at electron beam energy 55 eV and density $1.0 \times 10^{10}/cm^3$. Lines near visible range (400-700 nm) and lines with relatively big strength are shown in this table.

| Upper level | Lower level | Wavelength/nm | Intensity |
|---|---|---|---|
| $(4f^{14}5s^25p^65d^1)_{5/2}$ | $(4f^{14}5s^25p^65d^1)_{3/2}$ | 1124.20 | 3.1 |
| $(4f^{13}5s^25p^65d^2)_{7/2}$ | $(4f^{14}5s^25p^66d^1)_{5/2}$ | 215.70 | 2.0 |
| $(4f^{13}5s^25p^65d^2)_{7/2}$ | $(4f^{14}5s^25p^66d^1)_{5/2}$ | 198.13 | 1.5 |
| $(4f^{14}5s^25p^66p^1)_{1/2}$ | $(4f^{14}5s^25p^66s^1)_{1/2}$ | 152.88 | 2.3 |
| $(4f^{14}5s^25p^66d^1)_{5/2}$ | $(4f^{14}5s^25p^66p^1)_{3/2}$ | 97.48 | 12.9 |
| $(4f^{14}5s^25p^66s^1)_{1/2}$ | $(4f^{14}5s^25p^65d^1)_{5/2}$ | 87.32 | 10.3 |
| $(4f^{14}5s^25p^66p^1)_{1/2}$ | $(4f^{14}5s^25p^65d^1)_{3/2}$ | 52.95 | 20.8 |
| $(4f^{14}5s^25p^66p^1)_{3/2}$ | $(4f^{14}5s^25p^65d^1)_{5/2}$ | 50.50 | 28.3 |
| $(4f^{14}5s^25p^65f^1)_{7/2}$ | $(4f^{14}5s^25p^65d^1)_{5/2}$ | 34.02 | 19.9 |
| $(4f^{14}5s^25p^65f^1)_{5/2}$ | $(4f^{14}5s^25p^65d^1)_{3/2}$ | 33.08 | 12.3 |
| $(4f^{14}5s^25p^55d^2)_{7/2}$ | $(4f^{14}5s^25p^65d^1)_{5/2}$ | 24.79 | 20.6 |
| $(4f^{14}5s^25p^55d^2)_{5/2}$ | $(4f^{14}5s^25p^65d^1)_{3/2}$ | 24.37 | 13.8 |

The ground state of $W^{6+}$ is $4f^{14}5s^25p^6$ $^1S_0$, and there is no fine structure splitting. Several M1 transition lines near 500 nm with relatively large strengths belonging to the first excited configuration $4f^{14}5s^25p^55d^1$ are estimated by CRM, and listed in Table IV. Note that the simulated strengths of these lines are almost the same. However, no lines near 500 nm are observed in the present experiment.

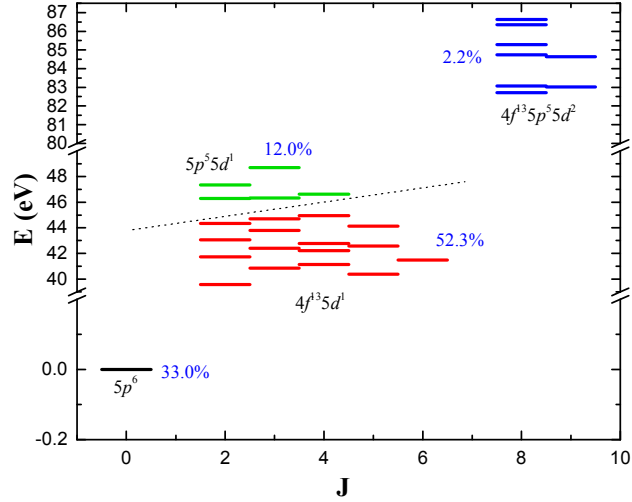

FIG. 3. A partial energy level diagram of $W^{6+}$ including energy levels with relatively high population. 1 energy level belonging to $4f^{13}5s^25p^6$ configuration with black color, 16 energy levels belonging to $4f^{13}5s^25p^65d^1$ configuration with red color, 5 energy levels belonging to $4f^{14}5s^25p^55d^1$ configuration with green color and 8 energy levels belonging to $4f^{13}5s^25p^55d^2$ configuration with blue color are shown. The total population of each configuration are marked with blue numbers, which are calculated by FAC at electron beam energy 70 eV and density $1.0 \times 10^{10}$/cm³.

Table IV. Results of CRM for $W^{6+}$ ions by FAC at electron beam energy 70 eV and density $1.0 \times 10^{10}$/cm³. Lines near visible range (400-700 nm) and lines with relatively big strength are shown in this table.

| Upper level | Lower level | Wavelength/nm | Intensity |
|---|---|---|---|
| $(4f^{13}5s^25p^65d^1)_4$ | $(4f^{13}5s^25p^65d^1)_5$ | 524.63 | 1.8 |
| $(4f^{13}5s^25p^65d^1)_4$ | $(4f^{13}5s^25p^65d^1)_5$ | 517.05 | 2.4 |
| $(4f^{14}5s^25p^55d^1)_3$ | $(4f^{14}5s^25p^55d^1)_2$ | 516.06 | 1.1 |
| $(4f^{13}5s^25p^65d^1)_3$ | $(4f^{13}5s^25p^65d^1)_4$ | 497.92 | 1.3 |
| $(4f^{13}5s^25p^65d^1)_5$ | $(4f^{13}5s^25p^65d^1)_6$ | 467.76 | 3.7 |
| $(4f^{13}5s^25p^65d^1)_3$ | $(4f^{13}5s^25p^65d^1)_4$ | 463.64 | 2.0 |
| $(4f^{13}5s^25p^65d^1)_1$ | $(4f^{14}5s^25p^6)_0$ | 29.99 | 14.0 |
| $(4f^{14}5s^25p^55d^1)_1$ | $(4f^{14}5s^25p^6)_0$ | 23.41 | 137.1 |
| $(4f^{13}5s^25p^55d^2)_7$ | $(4f^{13}5s^25p^65d^1)_6$ | 20.41 | 29.9 |
| $(4f^{13}5s^25p^55d^2)_6$ | $(4f^{13}5s^25p^65d^1)_6$ | 19.81 | 21.5 |
| $(4f^{14}5s^25p^55d^1)_1$ | $(4f^{14}5s^25p^6)_0$ | 18.95 | 385.3 |

Finally, theoretical wavelength, 548.61 nm, by the FAC code shows that the M1 transition in the ground configuration $4f^{13}5s^25p^6$ $^2F_{5/2} - {}^2F_{7/2}$ of $W^{7+}$ is the only strong transition in the visible range (see Fig. 4). This value is in consistency with that calculated by Berengut et al. (549.55 nm) [47]. However, Kramida et al. [48] evaluated this splitting to be 573.4 nm empirically from the measured $J = 5/2 - J = 7/2$ separation of $4f^{13}6s$, 7s, 6p, and 5f levels of $W^{6+}$. It is worth noting that this result is in excellent agreement with the present experimental value.

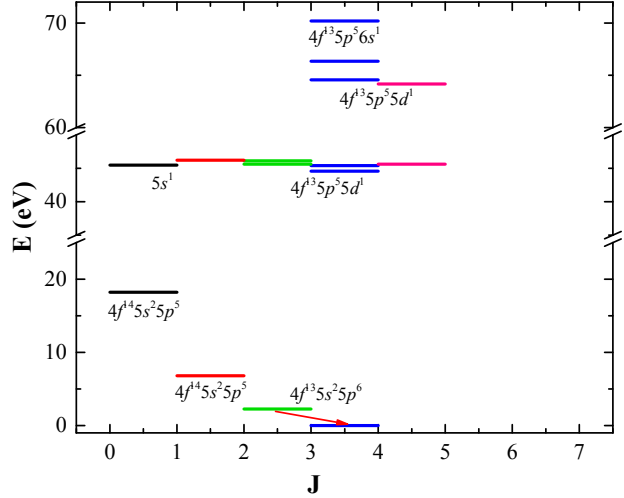

FIG. 4. A partial energy level diagram of $W^{7+}$ with the lowest, in energy, 15 energy levels from FAC calculations. The red arrow represents the M1 transition between the ground configuration $4f^{13}5s^25p^6$.

Regarding that the discrepancy in the wavelength between the FAC calculation (548.61 nm) and the experimental value (574.49 nm) is 4.50%, we have made a multi-configuration Dirac-Hartree-Fock (MCDHF) calculation by using the GRASP code [40, 41] in order to verify the source of this line.

In the MCDHF calculation, the active space approach is adopted to capture the main electron correlations. The correlation among the 5s, 5p and 4f valence electrons and the correlation between the 4s, 4p, 4d and n=3 in the core and the outer valence electrons are taken into account by the configuration state functions generated through restricted single (S) and double (D) excitations from the $3s^23p^63d^{10}4s^24p^64d^{10}4f^{13}5s^25p^6$ ground configuration to a virtual orbital set. The restriction means that only one out of n=3, 4s, 4p, and 4d core orbitals can be replaced by the virtual orbitals each time. The set of virtual orbitals are augmented layer by layer, and each layer is composed of orbitals with different angular symmetries up to 'g' except for the first layer where 'h' orbital is added as well. Four layers of virtual orbitals are required to make the fine-structure splitting converge. As can be seen from Table V, the fine-structure splitting of the $4f^{13}5s^25p^6$ ground configuration for $W^{7+}$ is not sensitive to the electron correlation. It is worth noting that the correlations related to the 3s, 3p, 3d, 4s, 4p and 4d core electrons are not negligible. They change the fine-structure splitting by around 1%. The Breit interaction and quantum electrodynamical effects (QED) are considered in the subsequent relativistic configuration interaction (RCI) computations. We found from Table V that the Breit interaction makes significant contribution to this fine-structure splitting, which reaches around 5%. The wavelength calculated by the GRASP code is in good agreement with our and other experimental values. This confirms that this line is corresponding to the M1 transition in the ground configuration of $W^{7+}$. For comparison, the present experimental and theoretical values of the wavelength for this line are listed in Table VI as well as other results available.

Table V. Fine-structure splitting (in cm$^{-1}$) and corresponding M1 transition wavelength (in nm) calculated by using multi-configuration Dirac-Hartree-Fock method. Breit and QED represent the Breit interaction and quantum electrodynamical effects, respectively.

| Models | Transition Energy (cm$^{-1}$) | Wavelength (nm) |
| --- | --- | --- |
| DF | 17899 | 558.69 |
| MCDHF | 18128 | 551.63 |
| Breit | 17425 | 573.89 |
| QED&Breit | 17435 | 573.56 |

Table VI. Comparison of experimental and theoretical results of M1 transition in ground term $4f^{13}5s^25p^6$ $^2F$ from $W^{7+}$.

| Name | Year | Type | Wavelength (nm) |
| --- | --- | --- | --- |
| Ryabtsev [49] | 2015 | Exp. | 574.46(16) |
| Mita [33] | 2016 | Exp. | 574.47(3) |
| This Work | 2018 | Exp. | 574.49(3) |
| Kramida [48] | 2009 | Theo. | 573.47 |
| Berengut [47] | 2009 | Theo. | 549.55 |
| This Work (by FAC) | 2018 | Theo. | 548.61 |
| This Work (by GRASP2K) | 2018 | Theo. | 573.56 |

The 574.49 nm line from $W^{7+}$ is observed at 59 eV (54.9 eV after correction) electron beam energy, which exceeds the ionization energy of $W^{4+}$, i.e. 51.6 eV, but lower than the ionization energy of $W^{5+}$, 64.77 eV and $W^{6+}$, 122.01 eV. This means that, the visible lines from $W^{7+}$ appear 2 charge states in advance in this experiment. Therefore, a hypothesis of indirect ionization in the charge-state evolution for generating $W^{7+}$ ions can be proposed, as shown in Fig. 5.

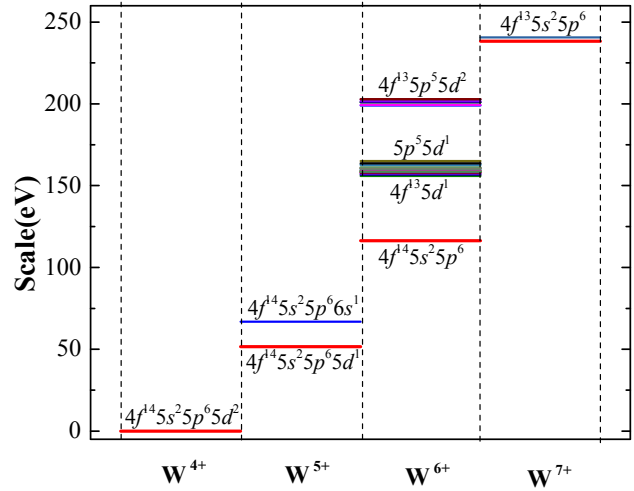

FIG. 5. Hypothesis of charge-state evolution of $W^{7+}$ from $W^{4+}$. The bold red lines represent the ground level of each charge state, while lines with other color represent the metastable level.

A large amount of $W^{5+}$ ions are produced through direct ionization from $W^{4+}$, when the electron beam energy exceeds the ionization energy 51.6 eV. For the $W^{5+}$ ion, it should be noted, that there exists a metastable state $4f^{14}5s^25p^66s^1$ $^2S_{1/2}$, 15.3 eV higher than the ground state $4f^{14}5s^25p^65d^1$ $^2D_{3/2}$, with the relatively high population. This leads to reduction of the ionization energy of $W^{5+}$ from 64.8 eV to 49.5 eV. Therefore, $W^{6+}$ ions could be yielded through indirect ionization from this metastable state at the same time once $W^{5+}$ ions occur.

According to the FAC calculation, as shown in Fig. 5, that there exists two metastable platforms for $W^{6+}$. The first metastable platform consists of two different configurations, that is, $4f^{13}5s^25p^65d^1$ and $4f^{14}5s^25p^55d^1$. Configuration $4f^{13}5s^25p^65d^1$ has 16 energy levels with 53.2% population in total and configuration $4f^{14}5s^25p^55d^1$ contains 5 energy levels with overall 12.0% population. The average energy of this platform is about 44.2 eV higher than the ground state, and less than the electron beam energy of 54.9 eV, so that the electrons could reach this platform by collision excitation. Moreover, the platform has extremely high population (up to 65%) and the long lifetime (on the millisecond order of magnitude). The adequate populations of these metastable states enable further collisional excitations from this platform towards higher energy levels.

The energy of the second platform of metastable states is approximately 40.5 eV higher than the first metastable platform (below the electron beam energy 54.9 eV), and includes several energy levels belonging to configuration $4f^{13}5s^25p^55d^2$. According to Pindzola et al. [50], the excitation cross-section for $5p$-$5d$ transition from configuration $4f^{13}5s^25p^65d^1$ of the first metastable platform (53.2% population) to configuration $4f^{13}5s^25p^55d^2$ of the second metastable platform is 214.96 Mb, which is much larger than other transitions. Such a large cross-section and the high population enhance the possibility for electrons to reach this platform by means of cascade excitation, and then get to the ground state of $W^{7+}$, whose energy is 37.3 eV higher. Consequently, $W^{7+}$ ions can be produced in this way.

In short, when the electron beam energy is tuned from 58 eV (51.1 eV after corrected) to 59 eV (54.9 eV after corrected), just exceeding the ionization energy of $W^{4+}$ (51.6 eV), $W^{5+}$ ions are generated in large amount by direct ionization. Then, $W^{6+}$ ions are produced through indirect ionization from the metastable state $4f^{14}5s^25p^66s^1$ $^2S_{1/2}$ of W$^{5+}$. In the same way, W$^{7+}$ ions are finally produced by indirectly ionization from the second metastable platform ($4f^{13}5s^25p^55d^2$) of W$^{6+}$. As a result, the M1 transition line from the W$^{7+}$ ground configuration, located near 574.49 nm, is observed. Energy levels, which play key roles in the indirect ionization process for W$^{7+}$ ions, are shown in Table VII.

Table VII. Information about energy levels taking effect in the indirect ionization process for W$^{7+}$ions. The energy here represents the relative energy compared to the ground state (0 eV) in each charge state.

| Charge state | Energy level | Energy (eV) | Population (%) | Lifetime (ms) |
|---|---|---|---|---|
| W$^{5+}$ | $(4f^{14}5s^25p^66s^1)_{1/2}$ | 15.30 | 0.50 | 0.02 |
| W$^{6+}$ First Metastable Platform | $(4f^{13}5s^25p^65d^1)_2$ | 39.57 | 2.22 | 0.44 |
| | $(4f^{13}5s^25p^65d^1)_5$ | 40.38 | 6.31 | 0.43 |
| | $(4f^{13}5s^25p^65d^1)_3$ | 40.83 | 3.20 | 0.44 |
| | $(4f^{13}5s^25p^65d^1)_4$ | 41.13 | 4.58 | 0.43 |
| | $(4f^{13}5s^25p^65d^1)_6$ | 41.48 | 7.40 | 0.44 |
| | $(4f^{13}5s^25p^65d^1)_2$ | 41.74 | 1.61 | 0.44 |
| | $(4f^{13}5s^25p^65d^1)_4$ | 42.21 | 3.97 | 0.43 |
| | $(4f^{13}5s^25p^65d^1)_3$ | 42.42 | 2.75 | 0.43 |
| | $(4f^{13}5s^25p^65d^1)_5$ | 42.58 | 5.21 | 0.43 |
| | $(4f^{13}5s^25p^65d^1)_4$ | 42.77 | 3.31 | 0.42 |
| | $(4f^{13}5s^25p^65d^1)_2$ | 43.06 | 1.23 | 0.42 |
| | $(4f^{13}5s^25p^65d^1)_3$ | 43.80 | 1.93 | 0.41 |
| | $(4f^{13}5s^25p^65d^1)_5$ | 44.13 | 3.15 | 0.41 |
| | $(4f^{13}5s^25p^65d^1)_2$ | 44.34 | 1.06 | 0.41 |
| | $(4f^{13}5s^25p^65d^1)_3$ | 44.70 | 1.87 | 0.43 |
| | $(4f^{13}5s^25p^65d^1)_4$ | 44.95 | 2.46 | 0.41 |
| | $(4f^{14}5s^25p^55d^1)_2$ | 46.29 | 1.53 | 1.03 |
| | $(4f^{14}5s^25p^55d^1)_3$ | 46.32 | 2.73 | 0.96 |
| | $(4f^{14}5s^25p^55d^1)_4$ | 46.63 | 3.75 | 1.06 |
| | $(4f^{14}5s^25p^55d^1)_2$ | 47.35 | 1.23 | 0.99 |
| | $(4f^{14}5s^25p^55d^1)_3$ | 48.69 | 2.76 | 1.04 |
| W$^{6+}$ Second Metastable Platform | $(4f^{13}5s^25p^55d^2)_8$ | 82.72 | 0.20 | 4.71 |
| | $(4f^{13}5s^25p^55d^2)_9$ | 83.02 | 0.52 | 4.90 |
| | $(4f^{13}5s^25p^55d^2)_8$ | 83.07 | 0.18 | 4.98 |
| | $(4f^{13}5s^25p^55d^2)_9$ | 84.64 | 0.52 | 4.03 |
| | $(4f^{13}5s^25p^55d^2)_8$ | 84.74 | 0.25 | 3.20 |
| | $(4f^{13}5s^25p^55d^2)_8$ | 85.28 | 0.26 | 4.20 |
| | $(4f^{13}5s^25p^55d^2)_8$ | 86.35 | 0.09 | 2.10 |
| | $(4f^{13}5s^25p^55d^2)_8$ | 86.62 | 0.12 | 2.39 |

## B. EUV spectra of $W^{7+}$

The spectra from the $W^{7+}$ ions in the EUV range from 17 nm to 26 nm are measured under nominal electron beam energy of 70, 73, 75, and 79 eV, respectively. The measurement time of the spectra is 2 hours and the beam current is kept constant at 3 mA. The results are shown in Fig. 6 and the correction of electron beam energy is shown in Table VIII.

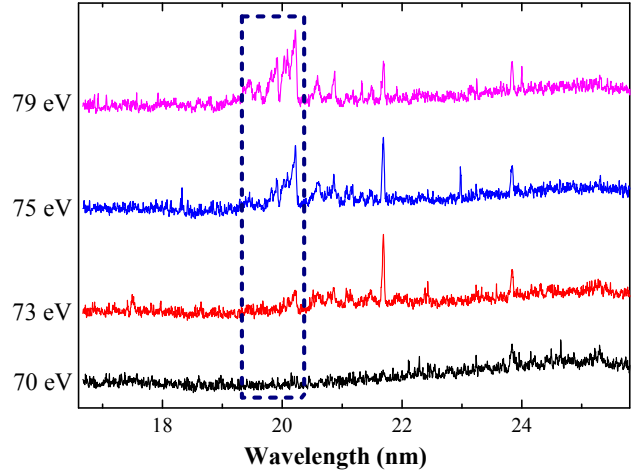

FIG. 6 Spectra of tungsten obtained at SH-HtscEBIT with the nominal electron beam energy 70,73,75 and 79 eV in EUV range 17-26 nm.

Table VIII. Correction of the electron beam energy when measuring spectra of EUV range: the set potential difference between the cathode and DT2 (nominal electron beam energy) $E_{set}$, the output potential difference between the cathode and DT2 $E_{out}$, the space charge effect from electrons and ions $V_{sp}$, and the finally corrected electron beam energy $E_{corr}$. The uncertainty for $V_{sp}$ and $E_{corr}$ are also given.

| $E_{set}$ (eV) | $E_{out}$ (eV) | $V_{sp}$ (eV) | $E_{corr}$ (eV) |
|---|---|---|---|
| 70.0 | 77.4 | 17.4±3.5 | 60.0±3.5 |
| 73.0 | 80.3 | 16.4±3.2 | 63.9±3.2 |
| 75.0 | 82.3 | 16.1±3.4 | 65.7±3.4 |
| 79.0 | 86.2 | 14.3±3.1 | 71.9±3.1 |

Different from spectra in visible range, lines in EUV domain are mostly in the form of transition arrays, and thus difficult to identify when the resolution of the spectrometer is not high enough. It can be seen from Fig. 1 and Fig. 6 that the lines at 19.3 to 20.3 nm (61.1 to 64.2 eV) do not appear at the same time as the visible line at 549.49 nm when the electron beam energy is 70 eV (60.0 eV corrected). After the electron beam energy reaches 73 eV (63.9 eV corrected), they emerge gradually. As the electron beam energy increases, the spectral lines in the transition array move toward lower wavelengths.

To explain this discrepancy between the visible and EUV spectra, RCI method in FAC is used. Totally 1127 energy levels are obtained by considering configuration involving $4f^{13}5s^25p^6$, $4f^{13}5s^25p^55d^1$, $4f^{13}5s^25p^55f^1$, $4f^{13}5s^25p^56s^1$, $4f^{13}5s^25p^56p^1$,

$4f^{13}5s^25p^56d^1$, $4f^{13}5s^15p^65d^1$, $4f^{13}5s^15p^65f^1$, $4f^{13}5s^15p^66s^1$, $4f^{13}5s^15p^66p^1$, $4f^{13}5s^15p^66d^1$, $4f^{14}5s^25p^5$, $4f^{14}5s^15p^6$, $4f^{12}5s^25p^65d^1$, $4f^{12}5s^25p^65f^1$, $4f^{12}5s^25p^66s^1$, $4f^{12}5s^25p^66p^1$, $4f^{12}5s^25p^66d^1$, $4f^{14}5s^25p^45d^1$, $4f^{14}5p^65d^1$ and $4f^{14}5s^15p^55d^1$. To identify these lines, spectra simulation is conducted by CRM under conditions of the electron energy 70 eV and density $1.0 \times 10^{11}/cm^3$. The results are shown in Fig. 7 as well as experimental results, and a good agreement can be found.

According to the theoretical results, the transition array at 19.3 to 20.3 nm mainly arises from the transitions between the higher excited state energy level (L209-L235) and the lower energy level (L0-L1). These include $5d$-$5p$, $5d$-$4f$ E1 transitions. The detailed energy level information is presented in Table IX.

The minimum electron beam energy, when lines in visible and EUV range of $W^{7+}$ ions occur, is 59 eV and 73 eV, respectively, which can be accounted for by the difference mechanism of spectral line production based on our FAC calculation. After the electron beam energy exceed the ionization energy of $W^{4+}$, $W^{7+}$ ions are generated by indirect ionization as mentioned above. As a result, the M1 line at 574.49 nm is observed at the 59 eV (54.9 eV after corrected) electron beam energy. However, the transition array at 20 nm comes from the E1 transitions from the higher-excited energy levels to the ground state of $W^{7+}$. Only if the corrected electron beam energy exceeds the excitation energy of upper levels, around 62 eV (see Table VIII and IX), can the direct collision excitation happen. Therefore, the transition array near 20.3 nm first appears as photon radiation from these excited states (see Fig. 6). As the electron beam energy further increases up to 65 eV, the higher excited levels are populated, giving rise to the appearance of transition array near 19.9 nm.

Table IX. Related energy level information. Here energy level represents the serial number in the calculated 1127 levels, and the energy represents the relative energy compared to the ground state (0 eV) in $W^{7+}$.

| Energy level | Configuration, J | Energy (eV) |
| --- | --- | --- |
| L0 | $(4f^{13}5s^25p^6)_{7/2}$ | 0.00 |
| L1 | $(4f^{13}5s^25p^6)_{5/2}$ | 2.26 |
| L209 | $(4f^{13}5s^25p^55d^1)_{5/2}$ | 61.46 |
| L210 | $(4f^{13}5s^25p^55d^1)_{7/2}$ | 61.65 |
| L212 | $(4f^{13}5s^25p^55d^1)_{9/2}$ | 61.86 |
| L214 | $(4f^{13}5s^25p^55d^1)_{5/2}$ | 62.06 |
| L215 | $(4f^{13}5s^25p^55d^1)_{9/2}$ | 62.07 |
| L216 | $(4f^{13}5s^25p^55d^1)_{7/2}$ | 62.30 |
| L218 | $(4f^{13}5s^25p^55d^1)_{7/2}$ | 62.62 |
| L222 | $(4f^{13}5s^25p^55d^1)_{9/2}$ | 62.93 |
| L223 | $(4f^{13}5s^25p^55d^1)_{5/2}$ | 63.14 |
| L228 | $(4f^{13}5s^25p^55d^1)_{7/2}$ | 63.69 |
| L230 | $(4f^{13}5s^25p^55d^1)_{5/2}$ | 63.92 |
| L233 | $(4f^{13}5s^25p^55d^1)_{7/2}$ | 64.57 |
| L235 | $(4f^{12}5s^25p^65d^1)_{3/2}$ | 64.92 |

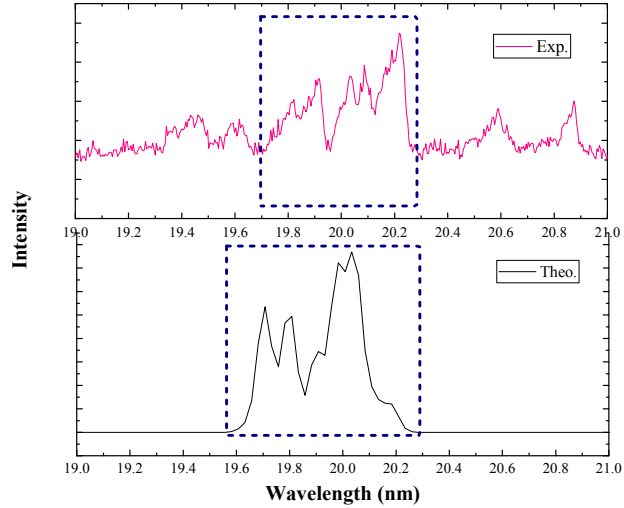

FIG. 7. Experimental and simulated spectra of $W^{7+}$ ions in EUV range 19-21 nm. The experimental spectra is obtained at nominal electron beam energy 79 eV while the simulated spectra is obtained by CRM at electron energy 70 eV and density $1.0 \times 10^{11}/cm^3$ with energy spread 3.5 eV.

**V. Conclusion**

The spectra of $W^{7+}$ are measured in the visible and EUV range at SH-HtscEBIT under extremely low electron beam energy conditions. The 574.49(3) nm M1 line of $W^{7+}$ is observed at the nominal electron beam energy of 59 eV which is below the ionization energy of $W^{6+}$. The multi-configuration Dirac-Hartree-Fock calculation further confirms the identification of this line. A hypothesis of charge-state evolution from $W^{5+}$ to $W^{7+}$ is proposed, based on our theoretical studies on the energy levels of these charge states, in order to explain the appearance of $W^{7+}$ spectra. Indirect ionization via cascade excitations from the long-lived metastable states of lower charge W ions play a key role in occurrence of $W^{7+}$. In addition, the EUV spectra at 75 eV as well as the FAC calculations also prove that $W^{7+}$ appears 2 charge states in advance according to the ionization energy.

**Acknowledgements**

This work was supported by the Chinese National Fusion Project for ITER No. 2015GB117000. One of the authors Q. Lu would like to thank Dr. A. Kramida from NIST for his helpful explanation on the atomic data of $W^{7+}$ in Ref. [48]. J.G. Li acknowledges support from NSFC (Grant No. 11874090).